\documentclass[sigconf, screen, dvipsnames, anonymous=false]{acmart}
\pdfoutput=1
\usepackage[ansinew]{inputenc}
\usepackage[show]{notes-alt}
\usepackage{booktabs}

\usepackage{graphicx}
\usepackage{multirow}

\usepackage{algorithm}
\usepackage{algpseudocode}

\usepackage{tikz}
\usepackage{tkz-graph}

\usepackage{pgfplots}

\usepackage{subcaption}

\title{Posthoc Interpretability of Learning to Rank Models using Secondary Training Data}

\author{Jaspreet Singh}
\affiliation{%
  \institution{L3S Research Center}
  \streetaddress{Applelstrasse 4, Hannover.}
}
\email{singh@l3s.de}

\author{Avishek Anand}
\affiliation{%
  \institution{L3S Research Center}
  \streetaddress{Applelstrasse 4, Hannover.}
}
\email{anand@l3s.de}

\acmConference[]{}{}{}

\begin{document}

\begin{abstract}

Predictive models are omnipresent in automated and assisted decision making scenarios. But for the most part they are used as black boxes which output a prediction without understanding partially or even completely how different features influence the model prediction avoiding algorithmic transparency. Rankings are ordering over items encoding implicit comparisons typically learned using a family of features using learning-to-rank models. In this paper we focus on how best we can understand the decisions made by a ranker in a post-hoc model agnostic manner. We operate on the notion of interpretability based on explainability of rankings over an interpretable feature space. Furthermore we train a tree based model (inherently interpretable) using labels from the ranker, called \emph{secondary training data} to provide explanations. Consequently, we attempt to study how well does a subset of features, potentially interpretable, explain the full model under different training sizes and algorithms. We do experiments on the learning to rank datasets with 30k queries and report results that serve show in certain settings we can learn a faithful interpretable ranker.

\end{abstract}

\maketitle

\section{Introduction}
\label{sec:intro}
Predictive models are all pervasive with usage in search engines, recommender systems, health, legal and financial domains. But for the most part they are used as black boxes which output a prediction, score or rankings without understanding partially or even completely how different features influence the model prediction. In such cases when an algorithm prioritizes information to predict, classify or rank information; algorithmic transparency becomes an important feature to keep tabs on restricting discrimination and enhancing explainability based trust in the system.

Take for example the \emph{European Union's new General Data Protection Regulation} which will take effect in 2018 that restricts automated individual decision-making which \emph{significantly} affects users. The law intends to create a right to explanation, whereby a user can ask for an explanation of an algorithmic decision that was made about them. The authors in~\cite{goodman2016european} highlight the opportunities for computer scientists to take the lead in designing algorithms and evaluation frameworks which avoid discrimination and enable explanation. 


Interpretability, traditionally, in machine learning has been studied under the ill-defined notion of understanding the mechanics of a learned models during decision making. Lipton in~\cite{lipton2016mythos} explores several possible notions of interpretability and tries to characterize it based on the \emph{desirable goals} and \emph{model properties}. In this paper, we make a preliminary attempt to answer the question: what can we infer and \emph{interpret} from an already trained model that helps us \emph{understand} its decisions on test data ,i.e., post-hoc interpretability. The notion of understanding is contingent on the goodness of an explanation, which we believe, is a function of the user interests, expertise and domain area. However in this paper we do not focus on the explanations themselves. Instead we focus on learning an interpretable model from a pretrained ranking model given we have a reasonable domain understanding and user expertise


To this extent, we initiate a study on what can one expect to learn or interpret from an already learned base model using features that are perceived to be interpretable in being able to be mimic the base model. In particular, we focus on interpreting models that learn rankings -- learning-to-rank-models that are currently widely employed in most information retrieval and recommender systems settings. We assume that base ranking models have been learned using training data that the interpreter is not privy to, but both the base ranker and interpreter operate in the same input space. Secondly, we assume that we have at our disposal a large number of test inputs. This is not unusual in practice where it is relatively easier to generate queries that produce rankings over a set of documents and feature computation is performed over only a limited top-k set of documents. With these test inputs we can recreate the ingredients needed to learn a new ranking model -- a train, validation and test set of query document pairs. The training data gathered by using the base ranker is referred to as \emph{secondary} training data in the remainder of this paper.  


We also assume reasonable, if not complete, system familiarity of the interpreter -- that is, the feature set considered by the interpreter is not arbitrary and is typically a subset of what is used by the base ranker. In our experiments, we chose a proper subset of features considered by the base model. Note that this is not always true in reality but a reasonable design choice to study this rather large experimental space. The choice of features by the interpreter is crucial in explaining the base ranking model and we note that explanations using certain features are more understandable than others. Content-based features are typically more understandable than using complex features derived from neural-networks that are themselves hard to interpret. For example, in explaining a ranking where $d_1 > d_2$ given a query term \texttt{brexit} based on the feature that computes query term presence in the page headline is more understandable. On the other hand, it is hard to explain comparisons if a feature computes the semantic similarity of the document with respect to a query in an embedding space. It is however debatable as to which content based features admit better explanations and we do not consider it as a part of this work. 


In this work, we attempt to learn a new interpretable model, that is not privy to the training regime of the original base model, which best resembles a given base ranker. Note that our notion of interpretability is purely based on how closely the interpreter resembles the rankings or output or the base ranker. We consider how an interpreter would learn from a large unconstrained set of secondary training examples that are generated by utilizing the predictions from the base rankers on our test input. We do not consider carefully constructed training examples that approximate only regions of the hypothesis space, i.e., are locally faithful. Instead, we consider \emph{global} models that train of large sets of random secondary training instances, not necessarily local, in-order to generalize to non-static unseen test data. Finally, we investigate the following research questions

\begin{itemize}
	
	\item \textbf{RQ I} Does an increasing amount of secondary training data help build better interpretable models ?
	\item \textbf{RQ II} Do the training algorithms used for base rankers affect the effectiveness of the interpreters ?
	\item \textbf{RQ III} How close can a global interpretable model get to mimicing the behavior base ranker ?
\end{itemize}

\vspace{-5pt}

\section{Related Work}
\label{sec:rel-work}

Algorithms prioritize information in a way that emphasizes or brings attention to certain things at the expense of others; by definition prioritization is about discrimination. As a result, there may be ramifications to individuals or other entities that should be considered during design. Search engines that are trained on a multitude of features with complex learning approaches are canonical examples. The criteria used in a ranking, how they are defined and datafed, and their weighting are essential design decisions that deserve careful consideration and scrutiny not just for performance but also in terms of interpretability.

\textbf{Interpretability of machine learned models:} There are two cases when considering the post-hoc interpretability of decisions made by the machine learned models -- the first, where we are aware of the procedure (linear models, neural nets or others) used to train the model and second where the model is used as a blackbox. In the latter model agnostic scenario, it is harder to acquire a deeper understanding of the model's behavior, and in particular how different features influence the model's predictions. The authors of~\cite{adler2016auditing} suggest an approach to identify features that might indirectly influence predictions via other, related features. While~\cite{adler2016auditing} tries to approximate the the global decision boundary, approaches like~\cite{ribeiro2016should} approximate local decision boundaries around a given prediction using only a set of interpretable features. In this work, we focus on approximating and understanding the global decision surface of a learned ranking model. Note that while our work seems to share similarities with adversarial learning~\cite{lowd2005adversarial}, we differ in two key aspects (i) our success is not measured relative to a cost model for the adversary and (ii) we want to approximate the global boundary rather than isolating susceptible areas of the boundary. Lipton in~\cite{lipton2016mythos} examines the motivations underlying earlier works on interpretability, finding them to be diverse and occasionally discordant. He then addresses model properties and techniques thought to confer interpretability, identifying transparency to humans and post-hoc explanations as competing notions. He discusses the feasibility and desirability of different notions, and question the oft-made assertions that linear models are interpretable and that deep neural networks are not.


\textbf{Interpretability of Learning-to-rank: } Learning to rank algorithms combine various document, query and behavioral features to induce an ordering amongst a set of documents retrieved for a given query. There are three broad types of learning to rank techniques -- pointwise, pairwse and listwise where each technique minimizes a different type of loss. While many algorithms have been suggested, only decision tree based appraoches have shown good performance and reasonable interpretability~\cite{ye2009stochastic}. There has been little work so far however in the IR community to address interpretability of blackbox learning-to-rank models. We propose to use a strategy followed by~\cite{adler2016auditing,lowd2005adversarial,ribeiro2016should} and other model agnostic approaches where we exploit the blackbox to generate labels to train a new interpretable model.



In this paper we make a first attempt to understand and intrepret a blackbox ranker.  In the next section, we detail our experiemtal procedure to learn from a blackbox ranker and then discuss key insights from our results.

\section{Experimental Setup}
\label{sec:experiments}
To best answer the research questions posed in Section~\ref{sec:intro}, we devised the following experiemtal setup:

All data used for the experiemts was derived from the learning-to-rank dataset~\cite{DBLP:journals/corr/QinL13} released by Microsoft. We use the dataset consisting of 30k web search queries. Each query has between 100 to 300 judged documents. The relevance judgments are obtained from a retired labeling set of a commercial web search engine (Microsoft Bing), which take 5 values from 0 (irrelevant) to 4 (perfectly relevant) and contains 132 features. 

In our experimental design we first create a base blackbox ranker using all the features $\mathcal{F}$ trained on a random set of 5k queries -- we call this model \textsf{M}. From the remaining queries, we then create 10 splits of increasing sizes -- $\{ 100, 200, 300, 400, 500,$ $1k, 2.5k, 5k, 7.5k, 15k\}$ for training the interpretable models. Note that split $i-1$ is a proper subset of split $i$. Instead of using the human judgements provided, the relevance labels for the training data are determined based on the ranking output of \textsf{M} -- called secondary training data. For our experiments we experimented with the following heuristic -- level 5 or highly relevant (ranks 1 to 5), level 4 or moderately relevant (ranks 6 to 10), and so on until irrelevant (label 0). Using the same heuristic we also create a set of 2.5k queries as validation data. We then finally selected 2.5k queries with the orginal labels as the test set. 

\subsection{Evaluation measures}
Rankings can be viewed as aggregations of preferences for intelligible explanations. A good interpretable ranker is one that preserves the preference of the base ranker. Standard IR metrics such as NDCG and Precision only serve as proxies for performance when we are interested in relative ordering of document pairs. To directly measure the effectiveness of our interpretable models, we consider how many relative orderings in pairs are preserved in the output ranking. Kendall's Tau considers the normalized difference between the number of concordant and discordant pairs when comparing two sets of rankings. We consider four evaluation measures to this extent -- Kendall's $\tau$, Kendall's $\tau@10$, NDCG@k and Precision@k. Additionally we also measure $\tau@10$ since users are often more interested in the top results. In our experiments we compare system variants against rankings produced by \textsf{M} for $\tau$ and $\tau@10$.

\subsection{System Variants}

We consider two variants of \textsf{M} -- a base ranker trained using (i) a list-wise learning approach (\textsf{M-L}) and (ii) a pairwise approach (\textsf{M-P}). Listwise learning-to-rank algorithms optimize an IR metric like NDCG directly whereas pairwise approaches minimze the number of discordant document pairs per query. For the incremental interpretable models we consider two scenarios -- (i) the system is aware of all features $\mathcal{F}$ used to train \textsf{M} and (ii) the system is only aware of a set of interpretable features $\mathcal{F'} \subseteq \mathcal{F}$. In both scenarios we report results when using a tree based learning-to-rank approach to train new rankers. We beleive that tree based models in conjunction with an interpretable feature space is the most interpretable in our setting. We use RankLib's implementation~\cite{dang2013ranklib} of LambdaMART (optimized for NDCG@10) and RankNet for the listwise and pairwise base rankers respectively. 

Ideally $\mathcal{F'}$ is directly dependent on the end user of such a system. In our experimental setup we assume the user is capable of interpreting simple content based features. $\mathcal{F'}$ is a set of 24 features that we believe are easiest to interpret for our user. We selected features ranging from simple boolean features representing the presence of query terms in the document to slightly more complex features like TF-IDF scores. In the future we intend to study the impact of $\mathcal{F'}$ as determined by user groups of varying expertise. A list of all 24 features can be found in the appendix.


\begin{table*}[ht!]
	\small
	\begin{tabular}{r@{\qquad}rrrr@{\hskip 6em}rrrr}
		\toprule
		\multirow{2}{*}{\raisebox{-\heavyrulewidth}{Training Size}} & \multicolumn{4}{c}{All Features (AM)} & \multicolumn{4}{c}{Interpretable Features (IM)} \\    \cmidrule{2-9} & NDCG@10 & Prec.@10 & $\tau$ & $\tau$@10 & NDCG@10 & Prec.@10 & $\tau$ & $\tau$@10 \\
		\midrule
		\textsc{100} & 0.3280 & 0.574 & 0.8664 & 0.7440
		&  0.2819 & 0.5482 &  0.4363 &  0.2004 \\
		\textsc{200} & 0.3397 & 0.5415 & 0.7328 & 0.6592
		& 0.2837 & 0.5507 &  0.4442 &  0.2692 \\
		\textsc{300} &0.3373 & 0.5932 & 0.7290 & 0.6736
		& 0.2859 & 0.5535 &  0.4400 &  0.2810\\
		\textsc{400} & 0.3396 & 0.594 & 0.7488 & 0.7108
		& 0.2849 & 0.5522 & 0.4632 & 0.3394\\
		\textsc{500} & 0.3398 & 0.5945 & 0.7197 & 0.7183
		& 0.2870 & 0.5535 &  0.3987 &  0.1806\\
		\midrule
		\textsc{1k} & 0.3397 & 0.5947 & 0.6960 & 0.6960
		&0.2893 & 0.5593 &  0.4341 &  0.2483\\
		\textsc{2.5k} & 0.3401 & 0.5941 & 0.6394 & 0.6489
		& 0.2886 & 0.5593 &  0.4088 &  0.1392\\
		\textsc{5k} & 0.3396 & 0.5928 & 0.7188 & 0.6709
		& 0.2879 & 0.5574 &  0.4216 &  0.1434 \\
		\textsc{7.5k} &0.3417 & 0.5948 & 0.7390 & 0.7144
		& 0.2877 & 0.5572 &  0.4127 &  0.2206\\
		\textsc{15k} & 0.3405 & 0.5936 & 0.7420 & 0.7173
		&0.2881 & 0.5571 &  0.4275 &  0.2180\\
		\midrule
		\textsc{M-P} & 0.3430 & 0.5569 &NA&NA
		&  0.3430 & 0.5569 &NA&NA\\
		
		\bottomrule
	\end{tabular}
	\caption{Models trained on increasing secondary training data from \textsc{M-P}}
	\label{tab:pairwise}
\end{table*}
\vspace{-5pt}

\begin{table*}
	\small
	\begin{tabular}{r@{\qquad}rrrr@{\hskip 6em}rrrr}
		\toprule
		\multirow{2}{*}{\raisebox{-\heavyrulewidth}{Training Size}} & \multicolumn{4}{c}{All Features (AM)} & \multicolumn{4}{c}{Interpretable Features (IM)} \\    \cmidrule{2-9} & NDCG@10 & Prec.@10 & $\tau$ & $\tau$@10 & NDCG@10 & Prec.@10 & $\tau$ & $\tau$@10 \\
		\midrule
		\textsc{100} & 0.3358 & 0.5794 & 0.6962 & 0.4039
		 & 0.2799	 & 0.542	 & 0.3939  & 0.0647 \\
		\textsc{200} & 0.3328	 & 0.5782	  & 0.6027	  & 0.3892
		 & 0.2825	 & 0.544	  & 0.4051  & 0.1277\\
		\textsc{300} & 0.3358	 & 0.5802	  & 0.6364	  & 0.3811
		& 0.2804	 & 0.5435	  & 0.3911	  & 0.1146\\
		\textsc{400} & 0.3366	 & 0.5819	  & 0.6763	  & 0.4030
		& 0.2779	 & 0.5414	  & 0.3990	  & 0.1383\\
		\textsc{500} & 0.3355	 & 0.5797	  & 0.6570	  & 0.3737
		& 0.2799	 & 0.5446	  & 0.3891	  & 0.1281\\
		\midrule
		\textsc{1k} & 0.3379	 & 0.5802	  & 0.6608	  & 0.4335
		& 0.2835	 & 0.5499	  & 0.2701	  & 0.1023\\
		\textsc{2.5k} & 0.3375	 & 0.5799	  & 0.6801	  & 0.4531
		& 0.2851	 & 0.5509	  & 0.2877	  & 0.1082\\
		\textsc{5k} & 0.3372	 & 0.5800	  & 0.7173	  & 0.4703
		& 0.2854	 & 0.5521	  & 0.4010	  & 0.1382\\
		\textsc{7.5k} & 0.3374	 & 0.5796	  & 0.7121	  & 0.4704
		& 0.2852	 & 0.5515	  & 0.4221	  & 0.1594 \\
		\textsc{15k} & 0.3377	 & 0.5804	  & 0.7368	  & 0.4952
		& 0.2857	 & 0.5514	  & 0.4400	  & 0.1522\\
		\midrule
		\textsc{M-L} & 0.4422 & 0.6439 &NA&NA
		&  0.4422 & 0.6439 &NA&NA\\
		
		\bottomrule
	\end{tabular}
	\caption{Models trained on increasing secondary training data from \textsc{M-L}}
	\label{tab:listwise}
\end{table*}
\vspace{-5pt}

\section{Results}

The first question we intend to answer is that if we used exactly the same features used for training \textsf{M} how close can we get to the rankings induced by \textsf{M}. Towards this, we first trained \textsf{M} on 5k queries, and trained models (called \textsf{AM}s -- All features Model) over secondary training data with all features $\mathcal{F}$. The results are reported in Tables~\ref{tab:listwise},\ref{tab:pairwise}. As expected the listwise \textsf{M} out-performs the pairwise variant significantly in both Precision and NDCG. We find that incrementally trained \textsf{AM}s do not improve significantly in either Precision or NDCG with increasing split size. The labels generated by \textsf{M} seemingly put a hard upper bound on performance irrespective of the type of learning approach used to train \textsf{M} and the number of training examples used. 

The kendall's $\tau$ results on the other hand are sensitive to the amount of training data used and the type of learning approach used for \textsf{M}. Somewhat surprisingly, when the base-ranker is \textsf{M-P}, we observe high correlation ($\tau$ $= 0.86$ and $\tau$@10$ = 0.74$) in the rank outputs after a 100 queries with performance steadily decreasing with increasing split size. This result is preliminary indication that a more careful approach to selecting training examples, akin to active learning~\cite{hanneke2014theory}, is required when \textsf{M} is pairwise learned. On the other hand, when \textsf{M} is listwise learned, we see steady improvement in the both $\tau$ and $\tau$@10 with increasing split size. However the $\tau$ and $\tau$@10 even after training on 15k queries is only $0.49$ and $0.74$ respectively. 

Our results show that even with all features available it is difficult to learn a faithful global reproduction of \textsf{M} just by using its' lables while increasing the number of training examples. Turning towards incrementally trained rankers that use only interpretable features $\mathcal{F'}$ (called \textsf{IM}s henceforth) we notice a similar trend to the \textsf{AM}s in NDCG and Precision. As expected the overall values are lower and vary insignificantly with increasing split size. When \textsf{M} is listwise learned, we once again see a steady increase in $\tau$ and $\tau$@10. The more training examples used for $IM$ the better it becomes at faithfully reproducing \textsf{M}'s ranking even though the NDCG and Precision remain unchanged. Although when using \textsf{M-P} we observe that the $\tau$ varies only slightly whereas $\tau$@10 is highly sensitive to the size of split. In our setup the split with 400 queries achieves the highest $\tau$@10 value of $0.33$ and nearly the highest Precision@10 ($0.5535$). This shows that the top 10 results have approximately the same number of relevant results and the ordering between the rank pairs is moderately correlated. Overall we find that when \textsf{M} is pairwise trained, \textsf{AM} and \textsf{IM} models can get closest to reproducing \textsf{M}'s results. Interestingly, \textsf{IM} also has the nearly the same precision as the base ranker. This is encouraging since most commercial search engines are trained on document pairs from click logs and the labeling mechanism we used to generate the secondary training data is inexpensive. Furthermore we can also easily cluster and isolate queries that the interpretable model can explain by measuring the $\tau$ and $\tau$@10 on the fly. 

In summary, for RQ I we see that increasing the amount of secondary training data is useful when considering \textsf{IM} or \textsf{AM} and \textsf{M-L}. We also found evidence that the type of base ranker has a significant on the performace of the interpretable ranker especially in terms of Precision and NDCG (RQ II). Finally for RQ III, we observe that under certain settings (split size 400 and \textsf{M-P}) we are able to get reasonable performance even with a subset of interpretable features. This allows us to provide faithful yet simple posthoc explanations of \textsf{M-L}.





\section{Conclusions \& Future Work}

Interpretability will be a key requirement in a future that is governed by algorithms. Learning-to-rank has been relatively unexplored in that regard and this paper tackles model agnostic post-hoc interpretability. We use the base blackbox ranker to produce secondary training data to learn a new interpretable model. The interpretable model uses a subset of the features used to train the base ranker. We defined an interpretable feature space consisting of 24 content based features. In our experiments we observed both pairwise and listwise learned base rankers using a dataset of 30k queries. We found that increasing the amount of secondary training data for the interpretable ranker leads to higher correlation with the base ranker when using a listwise learning algorithm. We also observed that we can faithfully reproduce the ranking of a pairwise trained base ranker even with a small amount of secondary training data. Standard IR metrics improve very slowly with the increasing amount of training data irrespective of the learning approach. Finally we found that content based features, assumed to be interpretable, perform poorly even with large quantities of secondary training data for the chosen dataset.

Once we develop better techniques to learn interpretable rankers, we envisage the following important scenarios that require the system to provide an explanation:
(i) \emph{Explain pairs -} Given a chosen pair of items from a ranked list of items, why is one item preferred or ranked higher than the other ?
(ii) \emph{Explain item vs rest (top or bottom) -} Given an item of interest, why is it placed in its current position. In other words, why are the items above it better and below it worse.
(iii) \emph{Explain top vs Bottom -} Given a ranking why are top-k items better than the remaining in the ranking. For example, why should I or should I not go the next page search results page.

In the future we would like to address the problem of active training data selection, selecting interpretable feature spaces and also experiment with various labeling mechanisms for secondary training data.

\section{ACKNOWLEDGMENT}
This work was carried out in the context of the ERC Grant (339233) ALEXANDRIA. This work was also supported by an Amazon Research Award.



	
	
\bibliographystyle{acm}
\bibliography{references}

\end{document}